\documentclass[floats,floatfix,showpacs,amssymb,prd,twocolumn,superscriptaddress,nofootinbib,nolongbibliography,reprint]{revtex4-2}

\usepackage{amssymb,amsmath,verbatim,mathtools,needspace,enumitem,etoolbox,graphicx,physics,microtype,afterpage,bigints,gensymb,tabularx,soul}

\usepackage[dvipsnames, usenames]{xcolor}
\definecolor{linkcolor}{rgb}{0.0,0.3,0.5}
\definecolor{dodgerblue}{HTML}{1E90FF}
\usepackage[unicode, colorlinks=true, linkcolor=linkcolor, citecolor=linkcolor, filecolor=linkcolor,urlcolor=linkcolor, pdfusetitle]{hyperref}
\usepackage[all]{hypcap}
\usepackage[T1]{fontenc}
\usepackage[utf8]{inputenc}
\usepackage{orcidlink}
\usepackage{soul}

\newcommand{\ssim}{\mathchar"5218\relax\,}

\makeatletter
\newcommand*{\balancecolsandclearpage}{\close@column@grid \cleardoublepage \twocolumngrid}
\makeatother

\newcommand\prlsec[1]{\vspace{2mm}\noindent {\bf \emph{#1}---}}

\newcommand{\bham}{\affiliation{School of Physics and Astronomy \& Institute for Gravitational Wave Astronomy, University of Birmingham, \\ Birmingham, B15 2TT, United Kingdom}}
\newcommand{\milan}{\affiliation{Dipartimento di Fisica ``G. Occhialini'', Universit\'a degli Studi di Milano-Bicocca, Piazza della Scienza 3, 20126 Milano, Italy}}
\newcommand{\infn}{\affiliation{INFN, Sezione di Milano-Bicocca, Piazza della Scienza 3, 20126 Milano, Italy}}

\begin{document}

\title{Catalog variance of testing general relativity with gravitational-wave data}

\author{Costantino Pacilio$\,$\orcidlink{0000-0002-8140-4992}}
\email{costantino.pacilio@unimib.it}

\milan \infn 

\author{Davide Gerosa$\,$\orcidlink{0000-0002-0933-3579}}

\milan \infn \bham

\author{Swetha Bhagwat$\,$\orcidlink{0000-0003-4700-5274}$\,$}

\bham


\date{\today}


\begin{abstract}
Combining multiple gravitational-wave observations allows for stringent tests of general relativity, targeting effects that would otherwise be undetectable using single-event analyses. We highlight how the finite size of the observed catalog induces a significant source of variance. If not appropriately accounted for, general relativity can be excluded with arbitrarily large credibility even if it is the underlying theory of gravity. This effect is generic and holds for arbitrarily large catalogs. Moreover, we show that it cannot be suppressed by selecting ``golden'' observations with large signal-to-noise ratios.
We present a mitigation strategy based on bootstrapping (i.e.~resampling with repetition)~that allows assigning uncertainties to one's credibility on the targeted test. We demonstrate our findings using both toy models and real gravitational-wave data. In particular, we quantify the impact of the catalog variance on the ringdown properties of black holes using the latest LIGO/Virgo catalog.
\end{abstract}

\maketitle
\prlsec{Introduction}
\label{sec:intro}
Gravitational-wave (GW) detections of binary compact objects allow for new tests of general relativity (GR) in the strong-field regime \cite{Baker:2014zba,LIGOScientific:2016lio} adding up to those performed with other experimental and astrophysical probes \cite{Berti:2015itd,Will:2014kxa}. Such tests are limited by the intrinsic challenges of modeling the strong-field dynamics in theories of gravity beyond GR \cite{Witek:2018dmd,Okounkova:2019zjf,Okounkova:2020rqw,East:2020hgw}, which prevents a directed, model-dependent search~\cite{Yunes:2016jcc}. In this regime, one primarily relies on testing the null hypothesis that GR is the underlying theory of gravity \cite{LIGOScientific:2020tif}.

At the individual-event level, tests of GR have been performed since the very first GW detection of binary black holes (BHs)~\cite{LIGOScientific:2016lio} and more stringent tests have since then been reported using the increasing number of detections during the first three LIGO/Virgo observing runs~\cite{LIGOScientific:2019fpa,LIGOScientific:2020tif,LIGOScientific:2021sio}. Combining multiple events is key to measuring effects that are otherwise undetectable using single sources.

Existing approaches can be categorized as (i) multiplication of the individual likelihoods~\cite{DelPozzo:2011pg,Ghosh:2016qgn}, (ii) multiplication of the individual Bayes factors~\cite{Gossan:2011ha,Agathos:2013upa,Meidam:2014jpa} and (iii) hierarchical inference \cite{Zimmerman:2019wzo,Isi:2019asy,Isi:2022cii}. Multiplication of the likelihoods assumes that deviations have the same values across all the events (e.g., constraints on the mass of the graviton) while multiplication of the Bayes factors assumes that deviations in multiple events are uncorrelated (e.g., constraints on additional BH hair) \cite{Zimmerman:2019wzo}. Both assumptions are unrealistic and Ref.~\cite{Isi:2019asy} first proposed hierarchical inference as a consistent way of combining observations, similarly to that of hierarchical Bayesian inference used in GW population studies~\cite{Adams:2012qw,Mandel:2018mve,Vitale:2020aaz}. In this context, the consistency of the data with GR can be quantified by standard metrics such as credible levels and Bayes factors.

Care must be exercised when interpreting the results of tests of GR, as they can lead to incorrect conclusions in the presence of unmodeled physics (e.g., environmental effects \cite{Barausse:2014tra,Berti:2022xfj}, eccentricity \cite{Saini:2022igm,Bhat:2022amc}), systematics in the waveform templates \cite{Moore:2021eok,Toubiana:2023cwr}, stealth biases \cite{Vallisneri:2013rc}, and overlapping signals \cite{Hu:2022bji}. In fact, one could also revert the argument and use tests of GR as a complementary method to identify the presence of systematics~\cite{Maggio:2022hre}.

In this Letter, we investigate an additional source of uncertainty when performing catalog tests of GR, namely the variance originating from the finite size of the catalog itself. We stress that, \textit{even if the null hypothesis is correct}, it could be excluded with arbitrarily large credibility from the posterior of the deviation parameters when combining multiple events. The issue would be mitigated if one were to repeat the experiment multiple times, as large deviations would only occur in relatively few repetitions. However, by definition, we are only going to have one catalog that contains all the observations. 

Crucially, our key message is that the catalog variance does not invalidate the use of null tests of GR, but it must be accounted for when interpreting the results. First, we show that using Bayes factors provides a more conservative evidence against violations of the null hypothesis than the corresponding credible intervals might suggest. Second, we design a mitigation strategy by assigning uncertainties to credible intervals and Bayes factors. Since one cannot use multiple realizations, we propose bootstrapping as a partial remedy \cite{2020sdmm.book.....I}. In a nutshell, from the original dataset $\boldsymbol{d}=\{d_{1},\dots,d_{N}\}$ one resamples a new dataset with the same size $\boldsymbol{d}^{\rm boot}=\{d^{\rm boot}_1,\dots,d^{\rm boot}_N\}$ allowing for repetitions. When resampling $\boldsymbol{d}$ with replacement, there are  $\binom{2N-1}{N}$ distinct combinations and the probability of obtaining the original dataset is as small as $N! / N^N$ \cite{oeis}. 
This mimics a set of repeated experiments to study the distribution of the chosen estimators (Bayes factors or credible intervals), which can then be used to extract summary statistics (e.g.~standard deviation, interquantile range), thus providing uncertainty estimates. A similar strategy consisting of downsampling the original catalog multiple times was employed in Ref.~\cite{Ghosh:2017gfp} to illustrate the variance in the inspiral-merger-ringdown consistency test of GR.

We focus on hierarchical tests of GR as introduced in Ref.~\cite{Isi:2019asy} as they represent the most general case. First, we perform numerical experiments to show that the catalog variance holds for arbitrarily large catalogs and it cannot be mitigated by selecting the observations based on their signal-to-noise ratio (SNR). Then, we demonstrate the impact on real GW data by reproducing and extending a flagship test of GR. In particular, we consider the so-called {\sc pSEOBNR} test~\cite{Ghosh:2021mrv} which targets deviations in the dominant frequency and damping time of the ringdown portion of the signal and was recently applied to the GWTC-3 catalog~\cite{LIGOScientific:2021sio}. We show that, while the hierarchical analysis of the damping time appears to exclude GR with high credibility, the corresponding Bayes factor prefers GR and the bootstrapped distributions have significant support in favor of the null hypothesis.

\prlsec{Hierarchical inference}
We are interested in testing the null hypothesis (i.e.~GR is the true theory) using a deviation parameter $x$, which is scaled such that it vanishes when the null hypothesis $\mathcal{H}_0$ is satisfied
\begin{equation}
    \label{eq:null:hypothesis}
    \mathcal{H}_0:\mathcal{H}\wedge\{x=0\}\,,
\end{equation}
where $\mathcal{H}$ is a broader hypothesis. If the null hypothesis $\mathcal{H}_0$ is inconsistent with the data, we expect deviations $x$ to spread away from $0$ following unknown patterns that are set by the system parameters and the nature of the deviations. 
GR tests are performed by applying hierarchical population inference \cite{Mandel:2018mve} to reconstruct the distribution of $x$ from the observed events $\boldsymbol{d}=\{d_1,\dots,d_N\}$. We model the distribution of $x$ as a normal distribution $\mathcal{N}$ with mean $\mu$ and variance $\sigma^2$
\begin{equation}
    \label{eq:p:pop}
    p_{\rm pop}(x|\mu,\sigma^2)=\mathcal{N}(x|\mu,\sigma^2)\,.
\end{equation}
In terms of these hyper-parameters, the null hypothesis maps to $\mu=\sigma^2=0$.
The posterior is given by 
\begin{equation}
    \label{eq:p:lambda}
    p(\mu,\sigma^2|\boldsymbol{d})\propto\mathcal{L}(\boldsymbol{d}|\mu,\sigma^2)\pi(\mu,\sigma^2)\,,
\end{equation}
where the hierarchical likelihood 
\begin{equation}
    \label{eq:l:lambda}
    \mathcal{L}(\boldsymbol{d}|\mu,\sigma^2) = \prod_{i=1}^N\int dx~\mathcal{L}(d_i|x)p_{\rm pop}(x|\mu,\sigma^2)
\end{equation}
can be expressed in terms of the likelihoods $\mathcal{L}(d_i|x)$ of the individual observations and $\pi(\mu,\sigma^2)$ models the prior. 

Equation~\eqref{eq:l:lambda} assumes that all observations are independent of each other which would be violated for, e.g., overlapping events. Reference~\cite{Hu:2022bji} estimates the fraction of overlapping binary BH events detected by next-generation ground-based GW detectors to be between ~5\% and ~35\% depending on the binary BH merger rate. If one adopts the method of joint parameter estimation \cite{Janquart:2022nyz}, the factors on the right-hand side of Eq.~\eqref{eq:l:lambda} corresponding to overlapping events are
replaced by joint likelihoods and joint population priors.
The expressions above do not include selection effects~\cite{Mandel:2018mve,Vitale:2020aaz} because in this context we do not wish to reconstruct the underlying distribution of $x$ but only constrain its value using the set of observed sources. 

The choice of the population \eqref{eq:p:pop} to describe the observed deviations might seem simplistic. However, in this context one is less interested in reconstructing the actual functional form of $p_{\rm pop}$ than in constraining it away from $\mu=\sigma^2=0$. Therefore, the ansatz \eqref{eq:p:pop} can suffice to the scope of detecting deviations from the null hypothesis, even if it is not faithful to their actual distribution. In particular, Refs.~\cite{Isi:2019asy,Isi:2022cii} showed that a Gaussian distribution can identify deviations from the null hypothesis even when these follow more complex patterns.

The consistency with the null hypothesis can be quantified using the quantile
\begin{equation}
    \label{eq:Q0}
    Q_0=\int_{p\geq p(0,0)} p(\mu,\sigma^2|\boldsymbol{d})~d\mu d\sigma^2\,,
\end{equation}
which is defined such that $Q_0=0$ ($Q_0=1$) indicate full consistency (full inconsistency).

The Bayes factor $\mathcal{B}$ in favor of the null hypothesis $\mathcal{H}_0$ over the broad hypothesis $\mathcal{H}$ can be estimated using the Savage-Dickey density ratio~\cite{Dickey},
\begin{equation}
\label{eq:savage:dickey}
    \mathcal{B} = \frac{p(\mu=0,\sigma^2=0|\mathbf{d})}{\pi(\mu=0,\sigma^2=0)}\equiv\frac{\mathcal{L}(\mathbf{d}|\mu=0,\sigma^2=0)}{\mathcal{Z}}
\end{equation}
where $\mathcal{L}$ is the hyper likelihood of Eq.~\eqref{eq:l:lambda} and $\mathcal{Z}=p(\mathbf{d})$ is the evidence of the data under $\mathcal{H}$. Bayes factors are often interpreted using Jeffreys' scale \cite{jeffreys1998theory}, where $\mathcal{B}\geq 10^2$ ($\mathcal{B}\leq 10^{-2}$) denotes ``decisive'' evidence in favor of (against) the null hypothesis.

From Eq.~(\ref{eq:savage:dickey}), the Bayes factor scales as $\mathcal{B}\propto\Delta$, where $\Delta$ is the prior volume: wide priors favor the null hypothesis, and vice versa tight priors favor the alternative hypothesis. This implies one can artificially increase the odds for either of the two competing models by restricting or enlarging the prior volume~\cite{Chua:2020oxn}. In the following, we fix this ambiguity by restricting the original prior volume to the $(1-p)$ posterior credible interval {along each axis}. For concreteness, the fraction of discarded posterior samples is set to  $p=1.973\times10^{-9}$, which corresponds to a $6$-$\sigma$ interval if these were Gaussian distributions. We then rescale $\mathcal{B}$ by the ratio $\Delta_{\rm new}/\Delta_{\rm old}$ of the restricted and original prior volumes. The rationale behind our choice is that $\Delta_{\rm new}$ is {just as large} to encompass the vast majority of the posterior support and therefore, the resulting Bayes factor constitutes a somewhat {conservative} estimate when testing GR. We denote the resulting Bayes factor as $\mathcal{B}_{\star}$ to distinguish it from the generic expression in Eq.~\eqref{eq:savage:dickey}, where the ambiguity in the prior volume is not fixed.

\prlsec{Catalog variance}
\label{sec:catalog:variance}
If the null hypothesis is correct, one would naively expect that the posterior for $\mu$ and $\sigma^2$ would become sharper around $\mu=\sigma^2=0$ as more events are added to the catalog; vice versa, it would peak away from zero if the null hypothesis is violated in nature. 

It is straightforward to check this expectation using a toy model where $x$ has Gaussian likelihoods for all the events
\begin{equation}
    \label{eq:l:x}
    \mathcal{L}(d_i|x)\propto\mathcal{N}(x|\mu_{{\rm obs},i},\sigma^2_{{\rm obs},i})
\end{equation}
and errors are homoscedastic, i.e., $\sigma_{{\rm obs},i}=\sigma_{\rm obs}={\rm const.}$ In the limit of large catalogs $N\gg1$, Eqs.~\eqref{eq:p:lambda} and \eqref{eq:l:lambda} reduce to~\cite{Isi:2022cii} 
\begin{equation}
    p(\mu,\sigma^2|\boldsymbol{d})\approx p(\mu|\boldsymbol{d})p(\sigma^2|\boldsymbol{d})\,,
\end{equation}
with
\begin{equation}
    \label{eq:p:mu}
    p(\mu|\boldsymbol{d})\propto\mathcal{N}\left(\mu\left|{\rm mean}(\mu_{\rm obs}),\frac{{\rm var}(\mu_{\rm obs})}{N}\right.\right)
\end{equation}
and
\begin{equation}
    \label{eq:p:sigma2}
 p(\sigma^2|\boldsymbol{d})\propto\mathcal{N}\left(\sigma^2\left|{\rm var}(\mu_{\rm obs})-\sigma_{\rm obs}^2,
    \frac{2\,{\rm var}(\mu_{\rm obs})^2}{N}\right.\right)\,\!.
\end{equation}
The true value of $x$ under the null hypothesis  is $x_{\rm true}=0$, which implies the $\mu_{{\rm obs},i}$'s are independently sampled from a normal distribution
\begin{equation}
    \label{eq:mu:sample}
    \mu_{{\rm obs},i}\sim\mathcal{N}(\mu_{{\rm obs},i}|\mu_{\rm true}=0,\sigma_{\rm obs}^2)\,,
\end{equation}
where the variance $\sigma_{\rm obs}^2$ accounts for the scatter due to noise in the detector consistently with the assumption of normal likelihoods \cite{Vallisneri:2007ev, Adams:2012qw}. The central limit theorem implies \begin{align}
    \label{eq:clt:1}
    {\rm mean}(\mu_{\rm obs})&\sim\mathcal{N}\left({\rm mean}(\mu_{\rm obs})\left|0,\frac{\sigma_{\rm obs}^2}{N}\right.\right)\,, \\
    \label{eq:clt:2}
    {\rm var}(\mu_{\rm obs})&\sim\mathcal{N}\left({\rm var}(\mu_{\rm obs})\left|\sigma_{\rm obs}^2,\frac{2~\sigma_{\rm obs}^2}{N}\right.\right)\,.
\end{align}
Plugging Eq.~\eqref{eq:clt:2} into Eq.~\eqref{eq:p:mu} shows that the $\mu$ posterior has variance $\ssim\sigma_{\rm obs}^2/N$ around ${\rm mean}(\mu_{\rm obs})$ at leading order in $N^{-1}$. By direct comparison with Eq.~\eqref{eq:clt:1}, it follows that $p(\mu|\boldsymbol{d})$ 
is not necessarily consistent with $\mu=0$ as $N$ increases, but there is a chance that the particular draw of $\{\mu_{{\rm obs},i}\}_{i=1}^N$ from \eqref{eq:mu:sample} shifts its peak away from the true value $\mu_{\rm true}= \sigma^2_{\rm true}=0$. A similar conclusion applies to the recovery of $\sigma^2$ by direct comparison of Eqs.~\eqref{eq:p:sigma2} and \eqref{eq:clt:2}. This toy model illustrates how the catalog variance is associated with the finite size $N$ of the catalog, along with a consistent inclusion of the scattering of measurements due to noise.

In writing Eqs.~\eqref{eq:p:mu} and \eqref{eq:p:sigma2} we neglected the effects of the prior $\pi(\mu,\sigma^2)$, i.e. we have assumed it is uniform and unbounded. However, the condition that $\sigma^2>0$ induces boundary effects in $p(\sigma^2|\boldsymbol{d})$. It follows that the posterior $p(\mu,\sigma^2|\boldsymbol{d})$ lacks a frequentist coverage of credible intervals, that is, it is not true that $Q_0> p$ in a fraction $(1-p)$ of similar experiments. The consequent difficulty of the statistical interpretation of $Q_0$ was already raised in Ref.~\cite{Chua:2020oxn}. 

We further highlight the impact of the catalog variance by considering $1000$ catalogs of $N=10^4$ events each, with Gaussian likelihoods for $x$ as per Eq.~\eqref{eq:l:x} and three different choices for the stochastic uncertainties. 
\begin{enumerate}[label=(\roman*)]
\item First, we consider the case of homoscedastic likelihoods with $\sigma_{\rm obs}=0.1$. 
\item Then, we assume heteroscedastic Gaussian likelihoods with $\sigma_{{\rm obs},i}=1/{\rm SNR}_i$, where ${\rm SNR}\in [10,1000]$ is a random variable distributed according to $p({\rm SNR})\propto {\rm SNR}^{-4}$ to mimic the density of SNRs expected from realistic GW detections \cite{Schutz:2011tw}. 
\item Finally, we isolate from the same heteroscedastic catalogs only the events with ${\rm SNR}\geq50$, which mimics a scenario where one performs tests of GR only on a loud subset of the available GW catalog.
\end{enumerate}
For each case, we draw the maximum-likelihood estimators $\mu_{{\rm obs},i}$ from $\mathcal{N}(\mu_{{\rm obs},i}|\mu_{\rm true}=0,\sigma_{{\rm obs},i}^2)$ to capture noise scattering.

\begin{figure}[t!]
    \centering
    \includegraphics[width=\columnwidth]{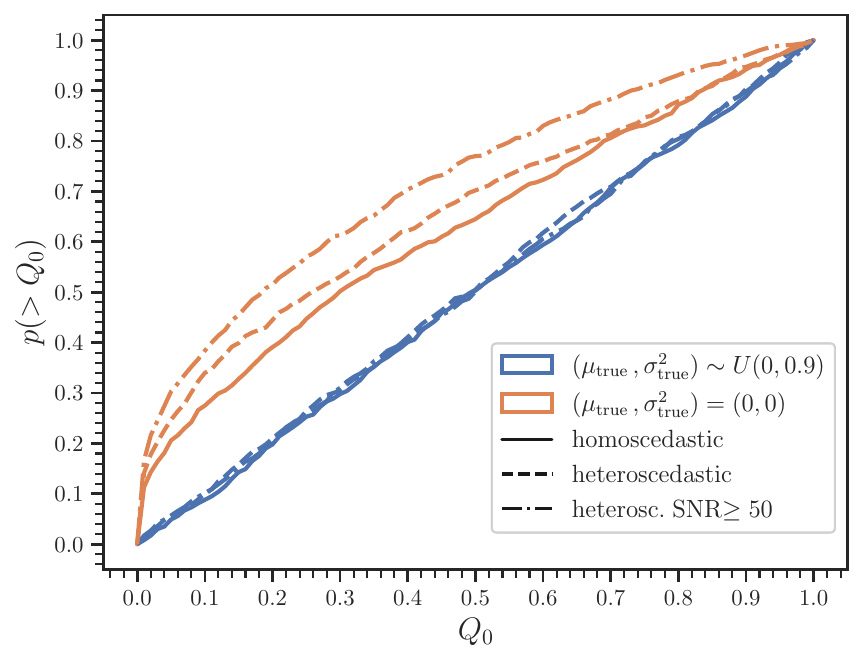}
    \caption{Cumulative distribution function of $Q_0$ for three sets of catalog realizations. We consider Gaussian likelihoods with homoscedastic errors (solid), heteroscedastic errors (dashed), and heteroscedastic with an SNR cut (dash-dotted). Orange and blue curves are produced by either assuming the null hypothesis $\mu_{\rm true}=0=\sigma_{\rm true}^2=0$ or drawing $(\mu_{\rm true},\sigma^2_{\rm true})$ from their prior, respectively.}
    \label{fig:pp:plot}
\end{figure}

We map the $(\mu,\sigma^2)$ posterior distribution using the \textsc{dynesty} implementation of nested sampling~\cite{Speagle:2019ivv} with 5000 live points and uniform priors over $\mu\in\mathcal{U}(-0.9,0.9)$ and $\sigma^2\in\mathcal{U}(0,0.9)$. While it is generally advised \cite{2020sdmm.book.....I} to use a log-uniform prior on scale parameters such as $\sigma^2$, Ref.~\cite{Gelman2006} shows on formal grounds that this causes issues in the context of hierarchical models if the likelihood has finite nonzero support for $\sigma^2=0$ (which in our case corresponds to the null hypothesis). Therefore, we opted for a uniform prior in~$\sigma^2$. 

Figure~\ref{fig:pp:plot} shows the resulting coverage of $Q_0$. To better highlight the peculiarity of the null hypothesis, we repeat the experiments without assuming that $\mu_{\rm true}=\sigma_{\rm true}^2=0$, but instead sample $(\mu_{\rm true},\sigma_{\rm true}^2)$ from their uniform priors at each catalog realization. As expected, when catalogs are drawn from uniform priors in $(\mu_{\rm true},\sigma_{\rm true}^2)$, the quantile $Q_0$ has a frequentist coverage. Instead, when drawing from the null hypothesis, the recovered values of $\sigma^2$ lie close to the edge of the prior, which produces an excess of low values of $Q_0$, thus pushing its cumulative distribution to the upper-right portion of Fig.~\ref{fig:pp:plot} --- see also the corresponding discussion in Ref.~\cite{Ghosh:2017gfp} in the context of tests of GR. That said, while $Q_0$ lacks a frequentist interpretation, it nonetheless provides an upper-bound estimate of the false alarm rate because $Q_0>p$ in less than a fraction $(1-p)$ of the catalog realizations. Figure~\ref{fig:pp:plot} also shows that restricting to high-SNR events does not reduce the effect of the catalog variance, in agreement with our interpretation based on the finite size of the catalog affected by noise realizations.

\prlsec{Bootstrapping} 
The catalog variance can be mitigated by assigning uncertainties to the chosen estimator. We showcase this idea by selecting a homoscedastic catalog realization with a high null-hypotesis quantile $Q_0=0.98$. The corresponding Bayes factor is $\log_{10}\mathcal{B}_{\star}=-0.65$, indicating substantial but not decisive evidence against the null hypothesis. After resampling for 1000 catalogs via bootstrap, we find that $Q_0>0.77$ and $\log_{10}\mathcal{B}_{\star}=-0.76^{+1.37}_{-2.61}$  at $90\%$ credibility. In particular, $\log_{10}\mathcal{B}_{\star}>0$ in $23\%$ of the bootstrapped catalogs, which is a non-negligible fraction and would suggest great care in claiming that the measurement provides evidence against the null hypothesis. This toy model shows that our proposed strategy is robust in mitigating false positives, even for catalogs with large credible quantiles. 
\begin{figure}[t!]
    \centering
    \includegraphics[width=
   \columnwidth]{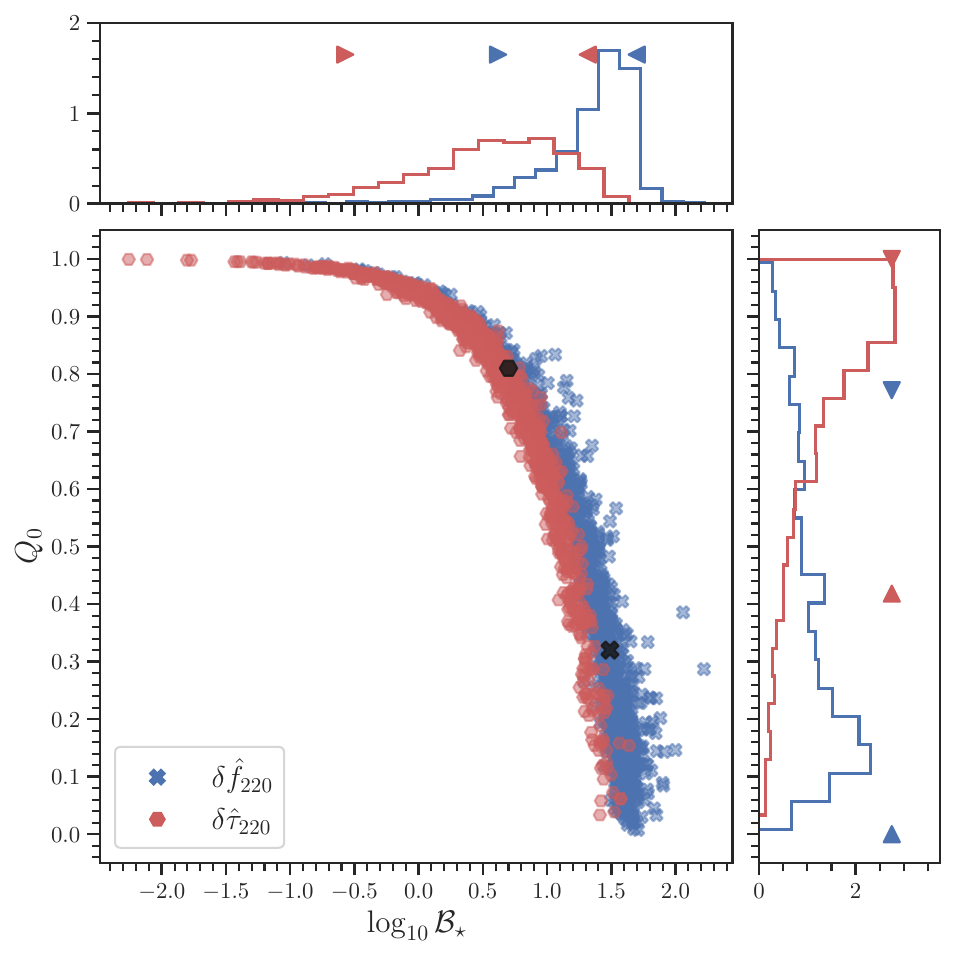}
    \caption{Distribution of Bayes factors $\log_{10}\mathcal{B}_{\star}$ and GR quantiles $Q_0$ over $1000$ bootstrapped realizations of the flagship \textsc{pSEOBNR} catalog test. Blue crosses and red hexagons indicate constraints on the ringdown frequency $\delta\hat f_{220}$ and damping time $\delta\hat \tau_{220}$, respectively. In the marginalized histograms, colored triangles indicate 90\% confidence intervals. Black markers indicate the values  of $\log_{10}\mathcal{B}_{\star}$ and $Q_0$ corresponding to the original catalog; this is just one possible realization among many!}
    \label{fig:scatter:plot}
\end{figure}

\prlsec{State-of-the-art application}
\label{sec:pseob}
We now apply our findings to a state-of-the-art test of strong-field gravity with GWs. We consider the \textsc{pSEOBNR} family~\cite{Brito:2018rfr,Ghosh:2021mrv,Maggio:2022hre,Toubiana:2023cwr} of binary-BH waveforms, which are obtained by augmenting effective-one-body templates with free parameters corresponding to fractional deviations in the quasinormal modes of the remnant BH. In the spirit of BH spectroscopy~\cite{Dreyer:2003bv,Berti:2005ys}, the \textsc{pSEOBNR} scheme has been used in tests of GR by allowing for deviations $\delta\hat f_{220}$ and $\delta\hat \tau_{220}$ in the dominant frequency and damping time respectively~\cite{LIGOScientific:2020tif,Ghosh:2021mrv,LIGOScientific:2021sio}. The latest iteration of these tests \cite{LIGOScientific:2021sio} uses 10 GW events and indicates a moderate deviation of $\delta\hat \tau_{220}$ from the GR value $\delta\hat \tau_{220}=0$. While insufficient to claim inconsistencies with GR, the authors themselves indicate this finding deserves further investigation.

In order to illustrate the role of the catalog variance in the interpretation of the results, we reproduce the \textsc{pSEOBNR} analysis of Ref.~\cite{LIGOScientific:2021sio} with a hierarchical combination of the events. For consistency with Ref.~\cite{LIGOScientific:2021sio}, we recover the $(\mu,\sigma)$ posterior and set uniform priors $\mu\in\mathcal{U}(-0.9,0.9)$ and $\sigma\in\mathcal{U}(0,0.9)$, covering a region that is  much broader than the resulting posterior. Quoting median and $90\%$ credibility, we obtain $\mu=0.02_{-0.04}^{+0.04}$, $\sigma<0.06$ for $\delta\hat f_{220}$ and $\mu=0.13^{+0.13}_{-0.11}$, $\sigma<0.19$ for $\delta\hat \tau_{220}$, which is in agreement with the analysis of Ref.~\cite{LIGOScientific:2021sio}. Using Eq.~\eqref{eq:Q0}, we quantify the consistency between GR and the data as $Q_0=0.32$ for $\delta\hat f_{220}$ and $Q_0=0.81$ for $\delta\hat \tau_{220}$. Using Bayes factors, we find $\log_{10}\mathcal{B}_{\star}=1.49$ for $\delta\hat f_{220}$ and $\log_{10}\mathcal{B}_{\star}=0.70$ for $\delta\hat \tau_{220}$. In particular, we note that in the case of $\delta\hat \tau_{220}$, even if $Q_0=0.81$, the log-Bayes factor is positive and hence it favors the null hypothesis.

We assign uncertainties to $Q_0$ and $\log_{10}\mathcal{B}_{\star}$ by generating 1000 bootstrapped catalog realizations. For each of these, we repeat the hierarchical analysis and extract the corresponding values of $Q_0$ and $\mathcal{B}_{\star}$. The analysis of Ref.~\cite{LIGOScientific:2021sio} uses 10 GW events, which implies there are $\ssim 10^5\gg 1000$ \cite{oeis} independent realizations and the probability of duplications is consequently small. Our results are shown in Fig.~\ref{fig:scatter:plot}. For $\delta\hat f_{220}$  we find that $\log_{10}\mathcal{B}_{\star}=1.45_{-0.83}^{+0.25}$ and $Q_0<0.77$ at 90\% confidence.  For $\delta\hat \tau_{220}$, we find
$\log_{10}\mathcal{B}_{\star}=0.62_{-1.19}^{+0.70}$ and $Q_0>0.42$.

Our bootstrap procedure returns broad histograms for $Q_0$; in particular, the credible quantile of the null hypothesis for $\delta\hat \tau_{220}$ can be as low as $Q_0=0.42$ within the $90\%$ range. Accounting for the catalog variance mitigates the significance of the inference performed with the original observed catalog. Moreover, the distribution of the Bayes factors for $\delta\hat \tau_{220}$ does not signal any substantial evidence against the null hypothesis at $90\%$ credibility; rather, $83\%$ of the samples have $\log_{10}\mathcal{B}_{\star}>0$, indicating support for the null hypothesis.

Finally, the correlation between $Q_0$ and $\log_{10}\mathcal{B}_{\star}$ shown in Fig.~\ref{fig:scatter:plot} indicates that Bayes factors provide weaker evidence against the null hypothesis than the corresponding credible level. In particular, while there are individual catalog realizations with $Q_0\approx1$, the corresponding Bayes factors barely meet the threshold for decisive evidence. 

\prlsec{Discussion}
\label{sec:discussion}
Combining information from multiple observations is a natural strategy to strengthen one's statistical inference on a physical phenomenon. Testing gravity with GWs is no exception. GR is a fundamental pillar of our understanding of the Universe and, when the stakes are so high, our confidence in experimental bounds becomes critical. The interpretation of tests of GR with GW catalogs depends on both the statistics (e.g.~quantiles and Bayes factors) as well as the techniques (e.g.~hierarchical inference) used to combine the inferences in favor or against the null hypothesis. Crucially, one must also quantify the catalog variance originating from the single  realization of the catalog of GW events at our disposal.

In particular, three key points are worth stressing:
\begin{enumerate}[label=(\roman*)]
  \item The net effect of accounting for the catalog variance is to soften one's claim in favor of violations of GR. 
  \item This is attained by ascribing uncertainties to point estimators of violations from GR.
  Uncertainties can be quantified by producing multiple mock catalogs. We propose a data-driven approach that does not rely on assuming a population of sources but instead resamples the observed catalog with repetition.
  \item The catalog variance does not vanish as either the size of the catalog or the SNR of the events increase (as long as they remain finite).
  \end{enumerate}

Points (i) and (ii) are best exemplified on the BH ringdown test we borrowed from the flagship  analysis of Ref.~\cite{LIGOScientific:2021sio}. We show that, while the current catalog presents a quantile $Q_0$ that might be interpreted as a moderate deviation from GR, this evidence turns out to be insignificant when the original measurement is considered as a part of a distribution of bootstrapped estimators. We have illustrated point (iii) with a toy model based on Gaussian likelihoods. 

Our findings lead to the conceptual issue of whether one should test the null hypothesis using Bayesian model selection in the context of tests of GR. As pointed out in Ref.~\cite{Chua:2020oxn}, reporting the evidence against GR with Bayesian estimators using free deviation parameters is questionable: results are  prior dependent and not reparametrization-invariant, while credible intervals lack a frequentist interpretation. On the other hand, a frequentist approach  based on the $p$-value only assesses the likelihood of the experimental outcome given the null hypothesis and can be considered more resilient. Unfortunately, implementing a pure $p$-value test in this context is, in practice, unfeasible because one would need to know the true population distribution of the events. 

Bootstrapping is a possible way out but only provides a partial solution. Bootstrap samples inevitably inherit the peculiarities (e.g.~outliers) of the specific catalog realization we have observed. 

A safer solution is to settle for weaker but more confident statements. This can be done trivially by breaking down the catalogs into chunks, using fewer events to compute the chosen estimator but obtaining multiple estimates; these estimates can then be used to construct histograms and credible quantiles for the estimator. While trivial, this strategy comes with the drawback that only a limited number of chunks can be obtained without sacrificing a significant fraction of the statistical power within the data. We speculate another promising avenue in this direction is to incorporate population inference into tests of GR \cite{Payne:2023kwj} while relying on the notion of ``Bayesian $p$-values''~\cite{Vallisneri:2023xay}.

While we concentrated on tests of GR, the catalog variance is a generic effect. For instance, astrophysical inferences from GW observations of binary populations~\cite{KAGRA:2021duu} and cosmological models \cite{LIGOScientific:2021aug} are impacted by the catalog variance in much the same fashion. The considerations put forward in this work are relevant to assess the statistical significance of some of those findings, especially when the significance itself is deemed to be weak.

\prlsec{Acknowledgments}
We thank Andrea Maselli, Riccardo Buscicchio, Golam Shaifullah, Nico Yunes, and the University of Illinois gravity group for discussions.
C.P. and D.G. are supported by ERC Starting Grant No.~945155--GWmining, Cariplo Foundation Grant No.~2021-0555, MUR PRIN Grant No.~2022-Z9X4XS, and the ICSC National Research Centre funded by NextGenerationEU. 
D.G. is supported by MSCA Fellowships No.~101064542--StochRewind and No.~101149270--ProtoBH, and Leverhulme Trust Grant No.~RPG-2019-350.
S.B. is supported by UKRI Stephen Hawking Fellowship No.~EP/W005727. 
Computational work was performed at CINECA with allocations through INFN and Bicocca.
\bibliography{testGR_cosmicvariance}
\end{document}